\advance\voffset by 1mm
\advance\hoffset by -7mm
\documentstyle[12pt]{article}
\def\ftoday{{\sl  \number\day \space\ifcase\month
\or Janvier\or F\'evrier\or Mars\or avril\or Mai
\or Juin\or Juillet\or Ao\^ut\or Septembre\or Octobre
\or Novembre \or D\'ecembre\fi
\space  \number\year}}
\newcommand{\journal}[4]{{\em #1~}#2\,(19#3)\,#4;}

\newcommand{\hpa}{\journal {Helv. Phys. Acta}}

\newcommand{\cmp}{\journal {Commun. Math. Phys.}}

\newcommand{\np}{\journal {Nucl. Phys.}}
\newcommand{\pl}{\journal {Phys. Lett.}}
\newcommand{\mpl}{\journal {Mod. Phys. Lett.}}

\makeatletter
\@addtoreset{equation}{section}
\makeatother

\newcommand{\es}{\\[3mm]}

\renewcommand{\d}{\delta}         \newcommand{\D}{\Delta}
\newcommand{\e}{\varepsilon}
\newcommand{\k}{\kappa}
        
\newcommand{\m}{\mu}
\newcommand{\n}{\nu}

\newcommand{\r}{\rho}

\newcommand{\vf}{{\varphi}}
\newcommand{\x}{\xi}

\renewcommand{\AA}{{\cal A}}
\newcommand{\BB}{{\cal B}}

\newcommand{\GG}{{\cal G}}

\newcommand{\VV}{{\cal V}}
\newcommand{\WW}{{\cal W}}

\newcommand{\Sla}{\raise.15ex\hbox{$/$}\kern -.70em D}

\newcommand{\lp}{\left(}\newcommand{\rp}{\right)}
\newcommand{\lc}{\left[}\newcommand{\rc}{\right]}
\newcommand{\lac}{\left\{}\newcommand{\rac}{\right\}}

\newcommand{\complex}{{\kern .1em {\raise .47ex
\hbox {$\scriptscriptstyle |$}}
    \kern -.4em {\rm C}}}
\newcommand{\real}{{{\rm I} \kern -.19em {\rm R}}}
\newcommand{\rational}{{\kern .1em {\raise .47ex
\hbox{$\scripscriptstyle |$}}
    \kern -.35em {\rm Q}}}
\renewcommand{\natural}{{\vrule height 1.6ex width
.05em depth 0ex \kern -.35em {\rm N}}}

\newcommand{\tr}{{\rm {Tr} \,}}

\newcommand{\pa}{\partial}

\newcommand{\fud}[2]{{\frac{\delta #1}{\delta #2}}}

\newcommand{\dfud}[2]{{\displaystyle{\frac{\delta #1}{\delta #2}}}}

\newcommand{\dint}{\displaystyle{\int}}

\newcommand{\ie}{{{\em i.e.},\ }}

\newcommand{\sla}{\raise.15ex\hbox{$/$}\kern -.57em}
\newcommand{\twiddle}{\lower.9ex\rlap{$\kern -.1em\scriptstyle\sim$}}

\newcommand{\equ}[1]{(\ref{#1})}
\newcommand{\eq}{\begin{equation}}
\newcommand{\eqn}[1]{\label{#1}\end{equation}}
\newcommand{\eea}{\end{eqnarray}}
\newcommand{\eqa}{\begin{eqnarray}}
\newcommand{\eqan}[1]{\label{#1}\end{eqnarray}}
\newcommand{\ba}{\begin{array}}
\newcommand{\ea}{\end{array}}
\newcommand{\eqac}{\begin{equation}\begin{array}{rcl}}
\newcommand{\eqacn}[1]{\end{array}\label{#1}\end{equation}}
\newcommand{\qq}{&\qquad &}
\begin{document}
\newcommand{\zc}{Z_{\rm c}}
\newcommand{\xt}{x^{\rm tr}}
\newcommand{\yt}{y^{\rm tr}}
\newcommand{\zt}{z^{\rm tr}}
\newcommand{\zb}{\bar{z}}
\newcommand{\kb}{\bar{\k}}
\newcommand{\pab}{\bar{\partial}}

\begin{flushright}
\begin{tabular}{l}
        UGVA---DPT 1994/01--843\\
        hep-th/9401140
\end{tabular}
\end{flushright}

\vspace{25mm}

\noindent{\Large{\bf Symmetries of the Chern-Simons Theory \\[2mm]
    in the Axial Gauge, Manifold with Boundary$^1$}}\vspace{10mm}

\noindent {\large S. Emery and  O. Piguet}\vspace{3mm}

\noindent D\'epartement de Physique Th\'eorique,
                                  Universit\'e de Gen\`eve\\
          24, quai Ernest Ansermet,
          CH -- 1211 Gen\`eve 4 (Switzerland)
\vspace{24mm}

\footnotetext[1]{Supported in part
                 by the Swiss National Science Foundation.}

\noindent {\small {\em Abstract.}
The field equations of the Chern-Simons theory quantized in
the axial gauge are shown to be completely determined
by supersymmetry Ward identities which
express the invariance of the theory under the topological
supersymmetry of Delduc, Gieres and Sorella together with
the usual Slavnov identity without requiring any action principle.
}\vspace{2mm}

\section{Introduction}
In a previous paper~\cite{2}, we show that all the Green
functions of the
Chern-Simon theory in three dimensions quantized in the
axial gauge can be completly and
uniquely determined by considering the usual BRS symmetry
together with the topological supersymmetry of
Delduc, Gieres and Sorella~\cite{dgs} without having to
invoque any action principle.
The construction goes as follow. The choice of a linear gauge condition
allows us to find the ghost equation by commuting this gauge condition
with the Slavnov identity. For the antighost equation, we show that it
is equivalent to the component of the supersymetry Ward identity along
the direction of the gauge defining vector. These two equations couple
only to the source of the Lagrange multiplier field, thus we can find a
recursion relation for the Green functions which involves only ghost,
antighost and Lagrange multiplier fields. For the gauge field we have no
such starting point. Nevertheless, using the transverse components of
the supersymmetry Ward identity, it is possible to connect all the Green
functions involving gauge and Lagrange multiplier field with the one's of
the previous set of fields.

In fact, there exists a deeper relation between this symmetry-based
approach and the content of an action principle defining the theory. Here
we will show that the gauge field equation is nothing but the consistency
condition between the transverse component of the supersymetry Ward identity
and the ghost field equation. Having this equation, one is able to find the
recursion relation which solves the gauge and Lagrange multiplier field sector
independently of the ghost-antighost sector. So in a certain sense, it put the
two sectors at the same level.

The paper is organized as follow. In spite of the fact that we want to
show that we can find the field equations without refering to any
action principle, we will begin with a short review of the 3-D
Chern-Simons theory quantized in the axial gauge in order to fix
the convention and notation. Section 3 is devoted to the study of
the symmetry of this theory and we will show, in section 4, how the
field equations arise from these symmetries.
Finally we will discuss in section 5 the case with boundary.

\section{Chern-Simons theory in the axial gauge}
The action of the Chern-Simons model in the axial gauge
reads\footnote{{\bf Conventions}: $\m,\n,\cdots = 1,2,3\ $,
$g_{\m\n}= {\rm diag}(1,-1,-1)\ $,
$\e^{\m\n\r}=\e_{\m\n\r}=\e^{[\m\n\r]}$,
$\e_{123}=1$.}
\eq\ba{rl}
\Sigma_{\rm CS}=&-\frac{1}{2}
\dint d^3x \epsilon^{\mu \nu \rho}\tr(A_\mu
\partial_\nu A_\rho +\frac{2}{3} g A_\mu A_\nu A_\rho)   \es
            &+\dint d^3x \tr(d n^\mu A_\mu+b n^\mu D_\mu c),
\ea\eqn{action}
with $D_\mu\cdot=\partial_\mu \cdot +g[A_\mu,\cdot]$ for the
covariant derivative. The gauge group is
chosen to be simple, all fields belong to the adjoint
representation and
are written as Lie algebra
matrices $\varphi(x)=\varphi^a(x) \tau_a$, with
\[
[\tau_a,\tau_b]= f_{ab}^c \tau_c,\qquad \tr(\tau_a
\tau_b)=\delta_{ab} .
\]
The canonical dimensions and ghost numbers of the fields
are given in
Table~\ref{dim-fp}.
\begin{table}[hbt]
\centering
\begin{tabular}
{|l|              r|     r|     r|    r|    } \hline
                &$A$   &$d$   &$b$  &$c$    \\ \hline
Dimension       &$1$   &$2$   &$2$  &$0$      \\ \hline
Ghost number    &$0$   &$0$  &$-1$  &$1$     \\ \hline
\end{tabular}
\caption[t1]{Dimensions and ghost numbers.}
\label{dim-fp}
\end{table}

The axial gauge is defined by the following
{\em gauge condition}
\eq
n^\m \dfud{\zc}{J^\m} + J_d =  0.
\eqn{gauge-cond}
Without loss of generality we can choose the vector $n$
defining the axial gauge as
\eq
(n^\m) = (0,0,1).
\eqn{gauge-vector}
The coordinates transverse to $n$ will be denoted by
\eq
\xt = (x^i,\ i=1,2).
\eqn{trans-coord}

\section{Symmetries and Ward identities}

The action \equ{action} is invariant~\cite{2} under the BRS
transformations $s$
\eq\ba{l}
sA_\m =- D_\mu c, \es
s b = d ,\es
s c = g c^2,\es
s d = 0,
\ea\eqn{brs}
as well as under the vector supersymetry $\n_\r$
given by
\eq\ba{l}
\n_\r  A_\mu =\epsilon_{\r  \mu \nu}n^\nu b, \es
\n_\r  b =0, \es
\n_\r  c = -A_\r,\es
\n_\r  d =\partial_\r b.
\ea\eqn{susy}

The BRS invariance of the theory can be expressed, formally, by the
functional identity
\eq
\tr\dint d^3x \left(
 - J^\m  [D_\m c]\cdot\zc   - g J_c [c^2]\cdot\zc
                - J_b \dfud{\zc}{J_d}   \right)
                       = 0.
\eqn{slavnov}
Here $\zc(J^\m,J_b,J_c,J_d)$ is the generating functional of
the connected Green
functions, $J^\m$, $J_d$, $J_b$ and $J_c$ denoting the
sources of the fields $A_\m$, $d$, $b$ and $c$, respectively.
We have used the notation
\[[O]\cdot\zc(J^\m,J_b,J_c,J_d)\]
for the generating functional of
the connected Green functions with the insertion of the
local field polynomial
operator $O$. Usually, such insertions must be renormalized,
their renormalization is controlled by coupling them to
external fields and the identity \equ{slavnov} becomes the
Slavnov identity~\cite{piguet-rouet}.
We shall however see below that, in the axial gauge,
these insertions are trivial and
thus the Slavnov identity is replaced by a local gauge Ward
identity.

The invariance under the supersymmetry transformations
$\n_\m$~\equ{susy} leads to the
{\em supersymmetry Ward identities}
\eq
\tr\dint d^3x \lp
 J^\m\e_{\r\m\n} n^\n\dfud{}{J_b}
+ J_c\dfud{}{J^\r} + J_d\pa_\r \dfud{}{J_b} \rp \zc
 = 0.
\eqn{susy-wi}

\section{Field equations recovered}

Now our goal is to recover the field equations using only the
functional identities~\equ{slavnov}, ~\equ{susy-wi} together
with the gauge condition ~\equ{gauge-cond}.

\subsection{Ghost-antighost sector}

It is well known that the choice of a linear
gauge condition leads to an equation for the ghost field. In the axial
gauge, it takes the form of the {\em ghost equation}
\eq
 - J_b  +
\lp n^\m\pa_\m\dfud{}{J_c} - g \lc J_d,\dfud{}{J_c} \rc
            \rp \zc = 0.
\eqn{ghost-eq}

Furthermore, we have already see ~\cite{2}
that the
projection of the supersymmetry Ward identity along the
gauge vector $n$:
$$
\tr\dint d^3x J_d \lp -J_c + n^\m\pa_\m \dfud{\zc}{J_b}\rp =0.
$$
leads to the local {\em  antighost equation}
\eq
 - J_c  +\lp n^\m\pa_\m\dfud{}{J_b}
            - g \lc J_d,\dfud{}{J_b} \rc \rp \zc = 0.
\eqn{antighost-eq}

The ghost equation \equ{ghost-eq}
and the antighost equation \equ{antighost-eq}
express the ''freedom'' of the ghosts in the axial
gauge~\cite{kummer}: they couple only to the $n$-component of the gauge
field, \ie to the external source $J_d$.
The effect of~\equ{ghost-eq} is to factorize out the contributions of the
ghost field $c$ to the composite fields appearing in the BRS
Ward identity \equ{slavnov}.  We can thus replace the
latter by the local {\em gauge Ward identity}:
\eq\ba{l}
-\pa_\m J^\m + \lp g\lc J^\m ,\dfud{}{J^\m}\rc
  + g\lc J_d, \dfud{}{J_d}\rc
    + g\lac J_b, \dfud{}{J_b}\rac \right.\es
\left.\qquad  + g\lac J_c, \dfud{}{J_c}\rac -
                   n^\m \pa_\m \dfud{}{J_d}\rp\zc = 0.
\ea\eqn{gauge-wi}

\subsection{Gauge sector}

Up to now, using only symmetry principle, we got the following set
of constraint (with the gauge vector $n_\m=(0,0,1)$)
\eqa
\GG^a(x) \zc&=&\lp\pa_3\fud{}{J_c}
      -g\lc J_d,\fud{}{J_c}\rc \rp^a\zc=J_b^a \nonumber\\
\AA^a(x) \zc&=&\lp\pa_3\fud{}{J_b}
      -g\lc J_d,\fud{}{J_b}\rc \rp^a\zc=J_c^a \nonumber\\
\WW^a(x) \zc&=&\lp\pa_3\fud{}{J_d}
      -g\sum_\vf\lc J_\vf,\fud{}{J_\vf}\rc \rp^a\zc=\pa_\m J_\m^a
                                               \nonumber\\
\VV_i(x)\zc&=& \tr \int d^3x\;
  \lp J^j\e_{ij}\dfud{}{J_b}
+ J_c\dfud{}{J^i} + J_d\pa_i \dfud{}{J_b} \rp \zc =0 \nonumber,
\eea
which correspond respectively to the ghost equation~\equ{ghost-eq}, the
antighost equation~\equ{antighost-eq}, the local gauge Ward
identity~\equ{gauge-wi} and the transverse component of the supersymmetry
Ward identity~\equ{susy-wi} written as functional differential equations.
These operators obey the following algebra
\eqa
\lc \WW^a(x),\WW^b(y)\rc&=&g\d^{(3)}(x-y)f^{abc}\WW^c(x) \nonumber\\
\lac\WW^a(x),\GG^b(y)\rac&=&g\d^{(3)}(x-y)f^{abc}\GG^c(x) \nonumber\\
\lac \WW^a(x),\AA^b(y)\rac&=&g\d^{(3)}(x-y)f^{abc}\AA^c(x) \nonumber
\eea
and
\eqa
\lc \WW^a(x),\VV_i(y)\rc&=&\d^{(3)}(x-y)\pa_3\pa_i\fud{}{J_b^a}
                             \label{comantigh}\\
\lac \GG^a(x),\VV_i(y)\rac&=&\d^{(3)}(x-y)\lp\pa_3\fud{}{J_i}
                                    \label{comgauge}
               -g\lc J_d,\fud{}{J_i}\rc\rp^a
\eea
all the other brackets being zero.
If we apply \equ{comantigh} and \equ{comgauge} on $\zc$, it gives the
following constistency conditions
\eqa
\pa_i\lp\pa_3\fud{}{J_c}
      -g\lc J_d,\fud{}{J_c}\rc \rp^a\zc&=&\pa_i J_b^a \label{der-antigh}\\
\lp\pa_3\fud{}{J_i}-g\lc J_d, \fud{}{J-i}\rc\rp^a \zc
                        &=&\e_{ij}J_j^a-\pa_i J_d^a. \label{gauge-eq}
\eea
{}~\equ{der-antigh} is nothing new, it is just the derivative of the
antighost equation, but~\equ{gauge-eq} corresponds exactly to the
{\em gauge field equation}.

We still have to deal with the dynamics of the Lagrange multiplier
field $d$. Here we cannot obtain its equation of motion by using the
functional identities given above. Nevertheless, gauge invariance implies
that Green's functions involving only the $d$ field are all zero. This can
be check from the Ward identity~\equ{gauge-wi} taken at
$J_\vf=0\ \forall \vf\not= d$.
Having noticed this, one is able to solve the full theory perturbatively
from \equ{ghost-eq}, \equ{antighost-eq} and
\equ{gauge-eq} and thus, to get the same
result as in \cite{2} without invoking any action principle.

\section{Manifold with boundary}
The case where the theory is defined on a manifold with bounbary is
important from the physical point of view. It is only in this case that
topological field theory  may pocess local observables, which then lie
on the boundary~\cite{wcmp121,hs-mplA691969}. In a previous
paper~\cite{emery-piguet}, we have already studied this case
using~\equ{action} and  shown that these observable are the two
dimensionnal conserved currents generating the Kac-Moody algebra~\cite{KM}
of the Wess-Zumino-Witten model~\cite{WZW}. Here we will show that the
alternative construction given above still holds in the presence of
boundary effects.

Let us now introduce as boundary the plane $\BB$ of equation
$x^3=0$. The effect of $\BB$ manifests itself as a breaking in the
Ward identity -- involving the locality and decoupling
conditions discussed in~\cite{emery-piguet} -- of the form
$\d(x^3)\D$ where $\D$ is some polynomial in the fields.
Their form is constrained by dimension and helicity
arguments. For the latter, it is convenient to choose the light-cone
coordinates for the transverse directions. At this point, one is faced
to the same problem of multiplying distributions at the same point as
in~\cite{emery-piguet}. A way to fix this ambiguity is to take
$$
\vf_\pm (\xt)=\lim_ {x^3\to\pm 0}\fud{\zc}{J_\vf (x)}
$$
for the insertion of the field $\vf(x)$ on the right (+) or on the
left (--) side of the boundary.

The functional identity which
generalizes the supersymmetry Ward identity~\equ{susy-wi} for the case
with boundary is
\eq\ba{rc}
\tr\dint d^3x \lp
 J^\m\e_{\r\m\n} n^\n\dfud{}{J_b}
+ J_c\dfud{}{J^\r} + J_d\pa_\r \dfud{}{J_b} \rp \zc=& \\[2mm]
  = \tr\dint d^2z \lp \k_{\r\pm}\dfud{\zc}{J^\r_\pm}
                             \dfud{\zc}{J_{b_\pm}}\rp &
\ea\eqn{susy-wi-bo}
where $\k_\r=(\k, \bar{\k},\x)$ are three {\sl a priori} independent
parameters of the breaking.

As for the case without boundary, the antighost equation is a direct
consequence of the projection of the supersymmetry Ward identity
along the gauge vector $n$. Thus, the antighost equation in presence
of the boundary takes the following form:
\eq
\lp \pa_3\dfud{}{J_b} - g \lc J_d,\dfud{}{J_b} \rc
        +\delta(x^3)\xi_\pm\dfud{}{J_b}\rp^a \zc= J^a_c  .
\eqn{bo-anti-eq}
For the ghost equation, we modify~\equ{ghost-eq} by adding a term
expressing the breaking due to the boundary. The parameter of this
breaking, which is {\em a priori} arbitrary, must be fixed to
$-\x_\pm$ due to consistency with \equ{bo-anti-eq}. So we get
\eq
\lp \pa_3\dfud{}{J_c} - g \lc J_d,\dfud{}{J_c} \rc
             -\delta(x^3)\xi_\pm\dfud{}{J_c}\rp^a \zc= J^a_b
\eqn{bo-gh-eq}
which is the ghost equation in presence of the boundary.\\
Then, the consistency between~\equ{bo-gh-eq} and~\equ{susy-wi-bo}
gives the gauge field equation for the case with boundary:
\eq
\lp \pa_3\fud{}{J^i}-g\lc J_d,\fud{}{J^i}\rc
     -\d(x^3)\lp\x-\k\rp_\pm\fud{}{J^i}\rp\zc=\e_{ij}J^j-\pa_iJ_d
\eqn{bo-gauge-eq}
where $\k=\bar{\k}$ due to consistency between the transverse
component of~\equ{susy-wi-bo}.

These equations
together with the fact that, due to BRS invariance, all the Green
functions involving only  $d$'s are zero, are sufficient to get
the results obtained in~\cite{emery-piguet}: the finiteness of the
theory as well as the existence of a Kac-Moody algebra on the boundary.
We can also check that all the field equations are invariant under
the discrete {\em parity transformation}
$z \leftrightarrow \zb\ ,\ u \rightarrow -u$,
under  which the fields transform as:
$$\ba{lclclclclcl}
A &\leftrightarrow& \bar{A}\ ,\qq A_u &\rightarrow& -A_u\ ,&&&&\\[2mm]
d &\rightarrow& -d\ ,\qq b &\rightarrow& -c\ ,\qq c &\rightarrow& b\ .
\ea
$$
As in~\cite{emery-piguet},
this invariance implies $\x_\pm=-\x_\mp$ and therefore, it allows
us to find the relation between the behaviour of the fields on the
boundary: $A$ and $\bar{A}$ can not be simultanuously zero, but
one of them does. Thus, if we choose $A(z,\zb,+0) =0$, then
$\bar{A}(z,\zb,+0) \neq 0$ generates the  Kac-Moody algebra.
For the other side, the roles of $A$ and $\bar{A}$ are interchanged:
$\bar{A}$ is zero on the boundary and $A$ generates the Kac-Moody
algebra.

\section{Conclusion}

We have shown that, given the field content of a topological field
theory, imposing BRS invariance and topological supersymmetry,
together with the choice of a linear gauge condition, is enough
to get all the field equations, except that of the Lagrange
multiplier, whose Green's functions are fixed by BRS symmetry. Thus,
at least for the specific theory treated in this paper, this is an
approach, for defining a theory, which is an alternative to the usual
one based on the action principle. This approach  seems to be easily
extendable to all topological fields theories.


\end{document}